\documentclass[english]{article}
\usepackage[affil-it]{authblk}

\usepackage[T1]{fontenc}
\usepackage[latin9]{inputenc}
\usepackage{refstyle}
\usepackage{color,hyperref}
    \catcode`\_=11\relax
    \newcommand\email[1]{\_email #1\q_nil}
    \def\_email#1@#2\q_nil{%
      \href{mailto:#1@#2}{{\emailfont #1\emailampersat #2}}
    }
    \newcommand\emailfont{\sffamily}
    \newcommand\emailampersat{{\color{red}\small@}}
    \catcode`\_=8\relax

\makeatletter

\RS@ifundefined{subref}
  {\def\RSsubtxt{section~}\newref{sub}{name = \RSsubtxt}}
  {}
\RS@ifundefined{thmref}
  {\def\RSthmtxt{theorem~}\newref{thm}{name = \RSthmtxt}}
  
  {}
\RS@ifundefined{lemref}
  {\def\RSlemtxt{lemma~}\newref{lem}{name = \RSlemtxt}}
  {}

\usepackage{amsmath}

\makeatother

\usepackage{babel}
\usepackage{mathrsfs}
\begin{document}
\title{Coulomb Law in the Non-Uniform Euler-Heisenberg Theory }
\author{A. D. Berm\'udez Manjarres}
\affil{\footnotesize Universidad Distrital Francisco Jose de Caldas\\ Cra 7 No. 40B-53, Bogot\'a, Colombia\\ \email{ad.bermudez168@uniandes.edu.co}}
\author{M. Nowakowski}
\affil{\footnotesize Departamento de F\'isica\\
Universidad de los Andes\\
Cra. 1E No. 18A-10
Bogot\'a, Colombia\\ \email{mnowakos@uniandes.edu.co}}
\author{D. Batic}
\affil{\footnotesize Department of Mathematics\\ Khalifa University of Science and
Technology, Main Campus, Abu Dhabi, United Arab Emirates \\ \email{davide.batic@ku.ac.ae}}

\maketitle
\begin{abstract}
We consider the non-linear classical field theory which results from
adding to the Maxwell's Lagrangian the contributions from the
weak-field Euler-Heisenberg Lagrangian and a non-uniform part which
involves derivatives of the electric and magnetic fields. We focus on 
the electrostatic case where the magnetic field is set to zero, and we
derive the modified Gauss law, resulting in a higher order differential
equation. This equation gives the electric field produced by stationary
charges in the higher order non-linear electrodynamics. Specializing for the case of a point charge,
we investigate the solutions of the modified Gauss law and calculate
the correction to the Coulomb law. 
\end{abstract}

\section{Introduction}
The search for new forces of short and long range or for new contributions to established
well known results has enjoyed over the  years a continued effort \cite{adelberger1, dobrescu, adelberger2}.
Most of the paradigms have been suggested and derived within quantum field theory and the famous ones like
the Casimir or van der Waals force have changed our understanding of quantum vacuum \cite{mostepanenko}.  

The established Coulomb law and the Newtonian gravitational potential are, of course, known to be  two
important pillars of physics. The quantum corrections to these results
are therefore of utmost relevance as they can have observable
consequences.
In the case of the Coulomb law these corrections are mostly, but as
discussed below not exclusively, derived by using the Fourier transforms
of Feynman amplitudes for elastic scattering giving rise
to vacuum polarization corrections (the Uehling-Serber potential)\cite{LL4}, fine and hyperfine
structure and finite size corrections \cite{one, two}. Its physical
relevance
gets revealed in precision atomic physics and even in nuclear tunneling
problems \cite{twelve}. In the case of gravity the procedure is
similar, but uses effective field theory approach in view of the lack
of a quantum gravity theory. Here corrections proportional to $\hbar$
have
been calculated and confirmed \cite{four, five}. They have applications
in quantum corrections to the Schwarzschild metric and cosmology
\cite{six, seven}.

Apart from the Fourier transform of the amplitude as  a tool to
calculate new potentials from quantum field theory (for yet different
examples see \cite{eight, nine}), there exists another source of
corrections in the case of the electromagnetic interaction. The light-light
scattering (recently observed directly \cite{lightlight}) gives rise
to a quantum corrected Lagrangian known as the Euler-Heisenberg theory
(see, e.g., \cite{LL4}). The resulting theory is, in nature,  a non-linear modification to Maxwell's equations subjected
to restrictions on the field strength, i.e. $E \ll m^2/e$,  (partly to avoid pair
production) and  its variation, $|\nabla E|/m \ll E$,  (partly due to mainly considering  uniform fields).
Since in the usual treatment derivatives of the field are ignored, we will call this case the uniform Euler-Heisenberg theory. To this,
corrections to the electromagnetic Lagrangian involving higher derivatives have been found in
\cite{mamaev, gusynin} and have been used in \cite{Roz, Shu, ludin1, ludin2}. We refer to these contributions as the non-uniform part
of the theory. Apart from dynamical issues \cite{ missingref2, ten, shock1, shock2, Rafelski}, the static case
of an electric potential deserves a special attention. In the purely
uniform Euler-Heisenberg theory the resulting equation for the
electric field produced by a static charge distribution reduces to a third order polynomial \cite{LL4, costa, kruglov},
and can be extended to higher order polynomials if we include higher
order uniform corrections \cite{eleven}. As shown below,  adding the non-uniform part
to the theory at first order gives a non-linear differential equation
which reduces to the polynomial one if higher order derivatives are
switched off.  We give the solutions for the asymptotic case where
the restrictions on the theory are satisfied.

\section{The Euler-Heisenberg Lagrangian and field equations}
The Euler-Heisenberg theory of non-linear electrodynamics derived from
quantum corrections to the Maxwell theory has been with us for more than
ninety years \cite{EHoriginal}. Many consequences of this theory
have been carefully investigated (see for example Ref \cite{EH2} and \cite{EH3}  for reviews and a large list of references), but in the majority of the cases
they focus in the modifications to the electromagnetic Lagrangian given by powers of $F_{\mu \nu}F^ {\mu \nu}$ and its
dual form $F_{\mu \nu}\tilde{F}^ {\mu \nu}$ while the terms involving derivatives of $F_{\mu \nu}$ are ignored. The above is justified since in many of the cases studied the uniform field approximation holds to a good degree (see \cite{ missingref2}  for a quantification of this statement). However, such derivative terms,
called here the non-uniform Euler-Heisenberg part, exist \cite{mamaev, missingref2}, and their consequences must be studied.
 
The full quantum mechanical corrections to the Maxwell Lagrangian are given
to the lowest order by

\begin{equation}
\mathscr{L}=\mathscr{L}_{0}+\mathscr{L}_{EH}+\mathscr{L}_{int}+b\left\{ \left(\partial_{\alpha}F_{\,\beta}^{\alpha}\right)\left(\partial_{\nu}F^{\nu\beta}\right)+F_{\alpha\beta}\left(\frac{\partial^{2}}{\partial t^{2}}-\nabla^{2}\right)F^{\alpha\beta}\right\} ,\label{FullL}
\end{equation}
where
\begin{equation}
\mathscr{L}_{0}=\frac{E^{2}-B^{2}}{8\pi}\end{equation}
is the Maxwell Lagrangian,
\begin{equation}
\mathscr{L}_{EH}=a\left[\left(E^{2}-B^{2}\right)^{2}+7(\mathbf{E}\cdot\mathbf{B})^2\right]\end{equation}
is the Euler-Kochel Lagrangian \cite{EK} (to which we will refer in the
text as
the uniform Euler-Hesienberg theory) and 
\begin{equation}
\mathscr{L}_{int}=A^{\mu}j_{\mu}\end{equation}
is the matter-field interaction term. 
For the constants we have $a=e^4/(360 \pi^2 m^4)$ 
 and $b=e^2/360 \pi m^2$ ($m$ being the electron mass and $e$ the charge). The non-uniform part
seems to be less suppressed with a lower power of $e/m$ which makes it interesting to search for
phenomenological consequences. In  Eq. (\ref{FullL})  the uniform and the non-uniform contributions to the effective Lagrangian are taken at the lowest possible order. Higher order terms exist \cite{ gusynin} and, in principle, could also be considered. However, such terms are suppressed by coefficients of higher powers in both  $\alpha$ and $1/m$. Our objective in this work is to derive the leading order asymptotic corrections to the Coulomb law. We  will argue later that for this purpose the Lagrangian (\ref{FullL}) is sufficient, hence, we will not take terms  of  higher order into account.

The equations of motion which
come from the variation of Lagrangian (\ref{FullL}) give corrections to
the classical Gauss and Ampere-Maxwell Laws. The Magnetic Gauss and
Faraday's laws do not proceed from any Lagrangian, therefore, they
remain the same in any non-linear version of electrodynamics. As a
consequence, the standard relation between fields and potentials remains
the same, a fact that will be used in the rest of this work. 

Specializing for the purely electrostatic case ($\mathbf{B}=0$ and time independent potentials),
we have $\mathscr{L}_{0}=\frac{1}{8\pi}E^{2}$, $\mathscr{L}_{EH}=aE^{4}$,
$\mathscr{L}_{int}=-\rho\phi$ and

\begin{align}
\left(\partial_{\alpha}F_{\,\beta}^{\alpha}\right)\left(\partial_{\nu}F^{\nu\beta}\right) & =\left(\nabla\cdot\mathbf{E}\right)^{2},\\
-F_{\alpha\beta}\nabla^{2}F^{\alpha\beta} & =2\mathbf{E}\cdot\left(\nabla^{2}\mathbf{E}\right).
\end{align}
Therefore, the Lagrangian (\ref{FullL}) simplifies to

\begin{equation}
\mathscr{L}=\frac{1}{8\pi}E^{2}+aE^{4}+b\left\{ \left(\nabla\cdot\mathbf{E}\right)^{2}+2\mathbf{E}\cdot\left(\nabla^{2}\mathbf{E}\right)\right\} -\rho\phi.
\end{equation}

Since in the electrostatic case we have $\mathbf{E}=-\nabla\phi$,
the Lagrangian under discussion contains higher order derivatives
of $\phi$. To obtain the equations of motion we cannot simply take
the standard Euler-Lagrange equations to first order. The classical
field theory of Lagrangians with higher derivatives can be found in
\cite{barut1,barut2}. Using the abbreviations $\partial_{ij}=\partial_i
\partial_k$ and $\partial_{ijk}=\partial_i \partial_j \partial_k$ the version of the Euler-Lagrange equations
we will need for our purpose is

\begin{equation}
\frac{\partial\mathscr{L}}{\partial\phi}-\partial_{i}\left(\frac{\partial\mathscr{L}}{\partial(\partial_{i}\phi)}\right)+\partial_{ij}\left(\frac{\partial\mathscr{L}}{\partial(\partial_{i}\partial_{j}\phi)}\right)-\partial_{ijk}\left(\frac{\partial\mathscr{L}}{\partial(\partial_{i}\partial_{j}\partial_{k}\phi)}\right)=0,\label{EL}
\end{equation}
where we have used the Einstein's summation convention and the indices
$i,j$ and $k$ run from 1 to 3\footnote{We point out that references \cite{ Shu, ludin1, ludin2} use the standard Euler-Lagrange equations, $\frac{\partial\mathscr{L}}{\partial\phi}-\partial_{i}\left(\frac{\partial\mathscr{L}}{\partial(\partial_{i}\phi)}\right)=0$,  to derive their correction to the Maxwell equations.  This is an incorrect procedure since we are dealing here with a Lagrangian with higher derivatives of the potentials.}. An important assumption we need for
the formula (\ref{EL}) to be valid is that $\phi$ has continuous
derivatives at least up to third order so that the Clairut theorem 
holds and
the order of the derivatives can be interchanged. 


Having the Lagrangian and the Euler-Lagrange equation to be used, we
now proceed to compute the equation for the electric field. We already
know that

\begin{align}
\frac{\partial\mathscr{L}_{int}}{\partial\phi} & =-\rho,\\
\partial_{i}\left(\frac{\partial(\mathscr{L}_{0}+\mathscr{L}_{EH})}{\partial(\partial_{i}\phi)}\right) & =-\nabla\cdot\left(\frac{1}{4\pi}\mathbf{E}+4aE^{2}\mathbf{E}\right).
\end{align}
The term $\left(\nabla\cdot\mathbf{E}\right)^{2}$ contains second
derivatives of the potential while $\mathbf{E}\cdot\left(\nabla^{2}\mathbf{E}\right)$ has
first and third derivatives. Plugging $\left(\nabla\cdot\mathbf{E}\right)^{2}$
into the Euler-Lagrange equation (\ref{EL}) we obtain the following

\begin{align}
\partial_{ij}\left(\frac{\partial\mathscr{L}}{\partial(\partial_{i}\partial_{j}\phi)}\right) & =-\partial_{ij}\left(\frac{\partial\mathscr{L}}{\partial(\partial_{i}E_{j})}\right)=-\partial_{ij}\left(\frac{\partial\left(\nabla\cdot\mathbf{E}\right)^{2}}{\partial(\partial_{i}E_{j})}\right) \nonumber \\
 & =-2\nabla^{2}\left(\nabla\cdot\mathbf{E}\right).
\end{align}
On the other hand, the first derivative variation of $\mathbf{E}\cdot\left(\nabla^{2}\mathbf{E}\right)$
gives

\begin{align}
\partial_{i}\left(2\frac{\partial[\mathbf{E}\cdot\left(\nabla^{2}\mathbf{E}\right)]}{\partial(\partial_{i}\phi)}\right) & =-\partial_{i}\left(2\frac{\partial[\mathbf{E}\cdot\left(\nabla^{2}\mathbf{E}\right)]}{\partial(E_{i})}\right)\nonumber \\
 & =-2\nabla\cdot\left(\nabla^{2}\mathbf{E}\right).
\end{align}
The third derivative variation of $\mathbf{E}\cdot\left(\nabla^{2}\mathbf{E}\right)$
can be restated as

\begin{equation}
\partial_{ijk}\left(2\frac{\partial[\mathbf{E}\cdot\left(\nabla^{2}\mathbf{E}\right)]}{\partial(\partial_{i}\partial_{j}\partial_{k}\phi)}\right)=-\partial_{ijk}\left(2\frac{\partial[\mathbf{E}\cdot\left(\nabla^{2}\mathbf{E}\right)]}{\partial(\partial_{i}\partial_{j}E_{k})}\right).\label{E3}
\end{equation}
Now, recalling that $\mathbf{E}\cdot\left(\nabla^{2}\mathbf{E}\right)=E_{k}\partial_{i}\partial_{i}E_{k}$, we obtain

\begin{equation}
\partial_{ijk}\left(2\frac{\partial[\mathbf{E}\cdot\left(\nabla^{2}\mathbf{E}\right)]}{\partial(\partial_{i}\partial_{j}E_{k})}\right)=-2\nabla^{2}\left(\nabla\cdot\mathbf{E}\right).
\end{equation}
Collecting the results we arrive at the modified Gauss law

\begin{equation}
b'\nabla^{2}\left(\nabla\cdot\mathbf{E}\right)+\nabla\cdot\left(\mathbf{E}+a'E^{2}\mathbf{E}\right)=4\pi\rho,\label{gauss1}
\end{equation}
where $a'=16\pi a$ and $b'=8\pi b$.

Equation (\ref{gauss1}) gives the electric field produced by any
arbitrary charge distribution once the quantum corrections from the effective Lagrangian are taken into account. Since this equation contains higher derivatives, special attention should be paid
to the boundary conditions. Fortunately,  the equation can be simplified by noticing
that the source $\rho$ is the same for both the Maxwell and the Euler-Heisenberg theories. Hence, we  can  replace $4\pi\rho$ by

\begin{equation}
4\pi\rho=\nabla\cdot\mathbf{E}_{c},\label{classical gauss}
\end{equation}
where $\mathbf{E}_{c}$ would be the field produced by the charge distribution due to the classical Gauss law. Thus, we can write (\ref{gauss1}) as

\begin{equation}
\nabla\cdot\left(b'\nabla^{2}\mathbf{E}\right)+\nabla\cdot\left(a'E^{2}\mathbf{E}+\mathbf{E}\right)=\nabla\cdot\mathbf{E}_{c},\label{gauss2}
\end{equation}
where we have used the fact that the potential is continuous up to
third order to interchange the order of the derivatives. 

As we are in the electrostatic case, we have that the electric field
must obey $\nabla\times\mathbf{E}=0$. Thus, we can cancel the divergence
operator on both sides of (\ref{gauss2}) to obtain the simplified
equation
\begin{equation}
b'\nabla^{2}\mathbf{E}+a'E^{2}\mathbf{E}+\mathbf{E}=\mathbf{E}_{c},
\end{equation}
which is of second order.
 
For the spherically symmetric case of a point charge we have
\begin{align}
\mathbf{E}_{c} & =\frac{e}{r^{2}}\widehat{\mathbf{r}},\\
\mathbf{E} & =E\,\widehat{\mathbf{r}},\label{E1}\\
E & =E(r).\label{E2}
\end{align}
In spherical coordinates, and taking into account the conditions (\ref{E1})
and (\ref{E2}), the vector Laplacian $\nabla^{2}\mathbf{E}$ reduces
to \cite{vectorlaplacian}

\begin{align}
\nabla^{2}\mathbf{E} & =\left(\nabla^{2}E-\frac{2E}{r^{2}}\right)\widehat{\mathbf{r}}\nonumber \\
 & =\left[\frac{1}{r^{2}}\frac{d}{dr}\left(r^{2}\frac{dE}{dr}\right)-\frac{2E}{r^{2}}\right]\widehat{\mathbf{r}}.
\end{align}
Care has to be taken since the Laplacian acting on a vector differs from its counterpart acting on a scalar. This 
well founded result goes back to the analogy between differential operators in curvilinear coordinates
and the covariant derivatives in differential geometry where the action of the covariant derivatives depends on
the the tensor type on which it acts \cite{hirota}.   

Finally, collecting everything together, we have the following equation  in spherical coordinates for the electric field produced by  a point charge

\begin{equation}\label{source}
b'\left[\frac{1}{r^{2}}\frac{d}{dr}\left(r^{2}\frac{dE}{dr}\right)-\frac{2E}{r^{2}}\right]+E+a'E^{3}=\frac{e}{r^{2}}.
\end{equation}

\section{The asymptotic solution}
In this section we investigate the behavior of the solutions to Eq.(\ref{source}). We focus in the asymptotic case of large $r$ since near the point charge we can expect the field to be too strong and too rapidly varying so $E \ll m^2/e$ and  $|\nabla E|/m \ll E$ most likely break down (though the inclusion of Euler-Heisenberg terms has the effect of taming the resulting produced field \cite{ ten}). For completeness, we give a short range solution to Eq. (\ref{source}) in an appendix at the end of the article.

To write an ansatz for an asymptotic solution $\left(mr\gg1\right)$ we notice that our differential equation contains the small parameters $a^{'}$ and
$b^{'}$. Therefore, it is clear that a solution can also be constructed by using these parameters. Secondly, the asymptotic
solution will have a form $E=e/r^2(1 + \xi)$ where dimensionless $\xi$ represents the quantum corrections of the form
$c_n/(rm)^n$. Since $a^{'}$ and $b^{'}$ contain inverse even powers of the mass $m$, only even exponents $n$ are allowed.
A quick dimensional analysis reveals that, for $n=4$ the coefficient $c_n$ must be proportional to $(b^{'})^2$ or $a^{'}$, 
for $n=6$ we would expect $a^{'}b^{'}$ or
$(b^{'})^3$ and finally for $n=8$ the coefficient is expected to be proportional to one of the terms of the form: $(a^{'})^2$,
$a^{'}(b^{'})^2$, $(b^{'})^4$.    
Therefore, the most general possible ansatz reads
\[
E(r)=\frac{e}{r^2}+a^{'}f_1(r)+b^{'}f_2(r)+(a^{'})^2 f_3(r)+(b^{'})^2 f_4(r)+a^{'}b^{'}f_5(r)+
\]
\begin{equation}\label{ans}
+(b^{'})^3f_6(r)+a^{'}(b^{'})^2f_7(r)+(b^{'})^4 f_8(r)\cdots,
\end{equation}
where $r>0$ and the $f_i$'s are differentiable functions in the radial variable $r$ yet to be determined. Substituting (\ref{ans}) into the differential equation (\ref{source}), combining like terms, i.e. powers of the parameters $a^{'}$ and $b^{'}$, and imposing that (\ref{ans}) is a solution of (\ref{source}) lead to the following system of ODEs for the unknown functions $f_i$: 
\begin{eqnarray}
f_1+\frac{e^3}{r^6}&=&0,\\
f_2&=&0,\\
f_3+3e^2\frac{f_1}{r^4}&=&0,\\
f_2^{''}+\frac{2f_2^{'}}{r}-\frac{2f_2}{r^2}+f_4&=&0,\\
f_1^{''}+\frac{2f_1^{'}}{r}-\frac{2f_1}{r^2}+\frac{3e^2 f_2}{r^4}+f_5&=&0,\\
f_2^{''}+\frac{2f_4^{'}}{r}-\frac{2f_4}{r^2}+f_6&=&0,\\
f_5^{''}+\frac{2f_5^{'}}{r}-\frac{2f_5}{r^2}+\frac{3e^2 f_2}{r^4}+\frac{3e f_2^2}{r^2}+f_7&=&0,\\
f_4^{''}+\frac{2f_6^{'}}{r}-\frac{2f_6}{r^2}+f_8&=&0,\\
&\vdots&.
\end{eqnarray}

Solving recursively the above system of equations leads to the following asymptotic expression for the field produced by a  point charge
\begin{equation}\label{pert}
E(r)=\frac{e}{r^2}-\frac{e^3}{r^6}a^{'}+\frac{3e^5}{r^{10}}(a^{'})^2+\frac{28e^3}{r^8}a^{'}b^{'}-\frac{1512e^3}{r^{10}}a^{'}(b^{'})^2+\cdots. 
\end{equation}
The result can also be given in terms of an expansion in $\alpha=e^2$ as
\begin{equation} \label{pert2}
E(r)=\frac{e}{r^2}\left(1-1.4\times10^{-2}\frac{\alpha^{3}}{(rm)^{4}}+8.8\times10^{-3}\frac{\alpha^{4}}{(rm)^{6}}+1.0\times10^{-2}\frac{\alpha^{5}}{(rm)^{8}}+...\right).
\end{equation}
 Equation (\ref{pert}) represents the Euler-Heisenberg contribution to the Coulomb law 
of an electron at large distances including contribution from the non-uniform terms.

Clearly, equation (\ref{pert}) reduces to Coulomb law when $b^{'}=a^{'}=0$, as it should be.  Additionally, we can check that equation (\ref{pert}) reproduces the known result from the uniform Euler-Heisenberg theory by noticing that in the case $b^{'}=0$ the differential equation (\ref{source}) reduces to a cubic equation whose only real solution is
\begin{equation} 
E_c(r)=\frac{2f(r)}{3a^{'}r}-\frac{r}{2f(r)},\quad f(r)=\sqrt[3]{\frac{(a^{'})^2 r}{16}\left(27e+3\sqrt{\frac{12r^4+81a^{'}e^2}{a^{'}}}\right)},
\end{equation}
which expanded around $a^{'}=0$ reproduces (\ref{pert}) with $b^{'}=0$. 
Hence, the new term proportional $a'b'r^ {-8}$ interpolates between $r^ {-6}$ and $r^ {-10}$ obtained 
from the Euler-Heisenberg theory without derivative corrections.

A second check consists in putting $a^{'}=0$ which sheds some light on the 
reason for why we do not encounter terms
proportional only to $b{'}^n$ in equation (\ref{pert}). 
In the case $b^{'}>0$ and $a^{'}=0$, two linearly independent solutions to equation (\ref{source}) are found to be
\begin{equation}
y_\pm(r)=\frac{e^{\pm i\frac{r}{\beta}}}{r^2}\sqrt{\mp i(r\pm i\beta)^2},\quad\beta=\sqrt{b^{'}}.
\end{equation}
Before taking the square roots appearing in the above expression, we need to fix a branch cut. In particular, we choose the principal value in such a way that the branch cut lies on the negative $r$-axis. Then,
\begin{equation}\label{theta}
\sqrt{\pm i}=e^{\pm i\frac{\pi}{4}},\quad\sqrt{(r\pm i\beta)^2}=\sqrt{r^2+\beta^2}e^{\pm i\vartheta},\quad
\vartheta=\arctan{\left(\frac{\beta}{r}\right)}
\end{equation}
and the particular solutions to (\ref{source}) can be cast into the compact form
\begin{equation}\label{complex}
y_\pm(r)=\frac{\sqrt{r^2+\beta^2}}{r^2}e^{\pm i\left(\frac{r}{\beta}-\frac{\pi}{4}+\vartheta\right)}.
\end{equation}
At this point, it is trivial to construct two linearly independent real solutions to equation (\ref{source}) by means of (\ref{complex}), and therefore, we limit us to state the general real and exact solution to the non-homogeneous ODE (\ref{source}) in the case $a^{'}=0$, namely
\begin{equation}\label{ex_sol}
E(r)=\frac{e}{r^2}+\frac{\sqrt{r^2+\beta^2}}{r^2} \left[c_1\cos{\left(\frac{r}{\beta}-\frac{\pi}{4}+\vartheta\right)}+c_2\sin{\left(\frac{r}{\beta}-\frac{\pi}{4}+\vartheta\right)}\right],
\end{equation}
where $\vartheta$ is a function of the radial variable given in (\ref{theta}).
Equation (\ref{ex_sol}) is not a physical result, but serves us to explain certain curious features. 
For large $r$ one would not even recover the correct Coulomb law from (\ref{ex_sol}) which actually means that the Coulomb law
comes from the uniform Euler-Heisenberg part of the theory.
Due to the presence of the term $1/\beta$ in (\ref{ex_sol}), we expect that in the case $b^{'}\ll 1$ and $a^{'}\neq 0$ the corresponding perturbative solution will not possess terms exhibiting powers of $b^{'}$ only as we discovered already in (\ref{pert}).

A few words of interpretation are in order. There is no one-to-one correspondence between the Lagrangian (\ref{FullL}) in coordinate space and the matrix elements in momentum space. Indeed, the term proportional to $a$ would result in momentum space in a point-like vertex of four photons and would not reflect the full light-light scattering. Similarly, the term proportional to $b$ in (\ref{FullL}) would correspond to corrections to the two-point function (vacuum polarization) and we should not expect to re-derive the Uehling-Serber potential. Indeed, Eq. (\ref{ex_sol}) shows that this is impossible and we should interpret the derivative terms as a non-uniform correction to the Euler-Heisenberg Lagrangian which taken by itself does not lead to physically viable results in the situation considered in this paper. Such an interpretation is reaffirmed by Eq (\ref{pert}) where $b$ appears only in combination with $a$. Notice that the Uehling-Serber potential does not have an inverse power law expansion for the case $mr\gg1$.

We can ask what is the effect of including higher order derivative
corrections in the Lagrangian. The next derivative term would be
proportional to $c=\frac{e^{4}}{m^{4}}$. As explained above, we expect
the effect of such a term to appear only in conjunction with $a$ , the
proportionality constant of the Euler-Heisenberg Lagrangian. Hence, by
dimensional analysis, we can expect the contribution of the next
higher derivative term to be of the order of
$ea'c'r^{-10}$. Therefore, such a new term would only modify the term
$r^{-10}$ in (\ref{pert}), the first contribution from the derivative
Lagrangian, $28ea'b'r^{-8}$, remains untouched. Similar lines of
arguments apply to other higher order terms appearing in
\cite{gusynin}. 

 Finally, we comment that we have limited ourselves to the 1-loop corrections to the Lagrangian. Incorporating a 2-loop correction to the first non-linear term in the Lagrangian \cite{2-loopa,2-loopb} would modify our $a$ by the addition of a term that is approximately $\alpha$ smaller. All of our results remain the same but with a small modification in $a$. This, and the effect of incorporating higher non-linear terms in the electric field produced by point (and an extended) charge, is discussed in \cite{eleven}.

\section{Conclusions}
Calculating quantum corrections to the Coulomb law has been an active field over decades. The interest goes back to the fact
that the Coulomb law governs many phenomena in classical and quantum physics. Therefore, knowing its exact quantum correction
is of some importance. Any new interaction or particle can be subjected to the test of reproducing well established results and as a 
consequence can, at the most, add corrections to these results. Similarly, any new higher order Lagrangian established within
the Standard Model should be examined carefully for its consequences.

The Euler-Heisenberg theory presents such an example. However, most of its effects have been derived using only
the uniform part neglecting hereby the non-uniform contribution with higher order derivatives. The dynamical and static properties
of the latter deserve a closer inspection. A first step towards such a program has been done in this paper: we have derived
a non-linear equation for the static electric field within the full Euler-Heisenberg theory (in the leading order).
We have solved this equation asymptotically and found terms which add to the standard solution derived using only
the uniform part. Although we have worked  with a point charge, the same methods also hold for the case of extended charge distribution just by taking the corresponding  classical field in the right hand side of Eq. (\ref{source}) in a 
manner similar to the procedure given in \cite{eleven}.

\appendix
\section{Appendix}
The uniform Euler-Heisenberg Lagrangian contains higher powers of the electric field $E$ 
(we restrict ourselves again to the static case with $B=0$) which can be used to derive higher order
polynomials for $E$ \cite{eleven}. To the lowest order a polynomial of the third order emerges which we
discussed in the main text. Solving these higher order polynomials displays a self-regulating effect:
the field strength of a proton can be brought below $m^2/e$ and the short distance solution can become viable. 
For the non-uniform part of the Euler-Heisenberg theory such a procedure does not exist, but could become
available in future. Therefore, we conclude the paper by deriving a power series solution of (\ref{source}) around the point $r=0$. We warn the reader that although the following solution is mathematically sound, we do not make much emphasis in its physical relevance due to the limitations in the range of validity of the Euler-Heisenberg theory.

We start by rewriting (\ref{source}) as follows
\begin{equation}\label{source1}
b^{'}r^2\frac{d^2 E}{dr^2}+2b^{'}r\frac{dE}{dr}-e+(r^2-2b^{'})E+a^{'}r^2 E^3=0.
\end{equation}
The above equation is a special case of the generalized second order Riccati equation treated in \cite{FL}, namely
\begin{equation}
(A_0+B_0 y)\frac{d^2 y}{dx^2}+(C_0+D_0 y)\frac{dy}{dx}-2B_0\left(\frac{dy}{dx}\right)^2+E_0+F_0y+G_0 y^2+H_0 y^3=0,
\end{equation}
where each coefficient can be Taylor expanded about $x=0$, more precisely
\begin{eqnarray}
A_0&=&x^2\sum_{k=0}^\infty a_k x^k,\quad B_0=x^2\sum_{k=0}^\infty b_k x^k,\quad C_0=x\sum_{k=0}^\infty c_k x^k,\\
D_0&=&x\sum_{k=0}^\infty d_k x^k,\quad E_0=\sum_{k=0}^\infty e_k x^k,\quad F_0=\sum_{k=0}^\infty f_k x^k,\\
G_0&=&x\sum_{k=0}^\infty g_k x^k,\quad H_0=x\sum_{k=0}^\infty h_k x^k,
\end{eqnarray}
with 
\begin{equation}\label{bed1}
e_0\neq 0\neq f_0, 
\end{equation}
and 
\begin{equation}\label{bed2}
A_0(0)=B_0(0)=C_0(0)=D_0(0)=G_0(0)=H_0(0)=0.
\end{equation}
In our case, both conditions (\ref{bed1}) and (\ref{bed2}) are satisfied provided that $b^{'}\neq 0\neq e$. Moreover, we have
\begin{equation}
A_0=b^{'}r^2,\quad C_0=2b^{'}r,\quad E_0=-e,\quad F_0=-2b^{'}+r^2,\quad H_0=a^{'}r^2.
\end{equation}
As in \cite{FL}, we are interested in a solution to (\ref{source1}) around $r=0$ admitting a power series of the form
\begin{equation}\label{ansatz}
E(r)=\alpha_0+\sum_{k=1}^\infty\beta_k r^k.
\end{equation}
The coefficients in (\ref{ansatz}) do not have in general the properties
\begin{equation}
\Delta_p=\mbox{det}\begin{bmatrix} 
    \alpha_0   & \beta_1 & \dots  &\beta_p    \\
    \beta_1    & \beta_2 & \dots  &\beta_{p+1} \\
    \vdots     & \vdots  & \ddots & \vdots\\
    \beta_p    & \beta_{p+1} &\dots &\beta_{2p}
    \end{bmatrix}\neq 0\quad\forall p=0,1,2,\cdots
\end{equation}
and
\begin{equation}
\Gamma_{2p+1}=\mbox{det}\begin{bmatrix} 
    \beta_1   & \beta_2 & \dots  &\beta_{p+1}    \\
    \beta_2    & \beta_3 & \dots  &\beta_{p+2} \\
    \vdots     & \vdots  & \ddots & \vdots\\
    \beta_{p+1}    & \beta_{p+2} &\dots &\beta_{2p+1}
    \end{bmatrix}\neq 0\quad\forall p=0,1,2,\cdots
\end{equation}
because as we will see soon in our case it turns out that $\Gamma_1=\beta_1=0$. Hence, according to \cite{wall}, the solution to equation (\ref{source1}) does not admit a representation in terms of a continued fraction of the form
\begin{equation}
E(r)=\cfrac{\alpha_0}{1+\cfrac{\alpha_1 r}{1+\cfrac{\alpha_2 r}{ 1+\ddots}}}.
\end{equation}
Finally, replacing (\ref{ansatz}) into (\ref{source1}) yields the following expansion for the electric field in a neighbourhood of $r=0$, namely
\begin{equation}
E(r)=-\frac{e}{2b^{'}}+\beta_2 r^2+\beta_4 r^4+\beta_6 r^6+\beta_8 r^8+\mathcal{O}(r^{10}),\label{Final}
\end{equation}
with 
\begin{eqnarray}
\beta_2&=&\frac{e[a^{'}e^2+4(b^{'})^2]}{512(b^{'})^4},\\
\beta_4&=&-\frac{e[3(a^{'})^2 e^4+16a^{'}(b^{'})^2 e^2+16(b^{'})^4]}{147456(b^{'})^7},\\
\beta_6&=&\frac{e[45(a^{'})^3 e^6+336(a^{'})^2(b^{'})^2 e^4+656a^{'}(b^{'})^4 e^2+128(b^{'})^6]}{188743680(b^{'})^{10}},
\end{eqnarray}
and
\begin{equation}
\beta_8=-\frac{e[585(a^{'})^4 e^8+5628(a^{'})^3(b^{'})^2 e^6+17232(a^{'})^2(b^{'})^4 e^4+16448a^{'}(b^{'})^6 e^2+512(b^{'})^8]}{211392921600(b^{'})^{13}}.
\end{equation}

It is peculiar to see that the electric field at the origin given by Eq. (\ref{Final}) has a negative sign. Near a positive charge Eq.(\ref{Final}) gives a field that points toward the charge.

\end{document}